\documentclass[prd,amsmath,amssymb,longbibliography,superscriptaddress,twocolumn,nofootinbib,10pt]{revtex4-1}

\pdfoutput=1

\usepackage{graphicx}
\usepackage{dcolumn}
\usepackage{bm}
\usepackage{amssymb}
\usepackage{latexsym}
\usepackage{booktabs}
\usepackage{amsmath}
\usepackage{multirow}
\usepackage{enumitem}

\usepackage[colorlinks=true, linkcolor=blue, citecolor=blue]{hyperref}

\begin{document}

\title{Observational challenges to holographic and Ricci dark energy paradigms: Insights from ACT DR6 and DESI DR2}

\author{Peng-Ju Wu}\thanks{Corresponding author}\email{wupengju@nxu.edu.cn}
\affiliation{School of Physics, Ningxia University, Yinchuan 750021, China}

\author{Tian-Nuo Li}
\affiliation{Department of Physics, College of Sciences, Northeastern University, Shenyang 110819, China}

\author{Guo-Hong Du}
\affiliation{Department of Physics, College of Sciences, Northeastern University, Shenyang 110819, China}

\author{Xin Zhang}\thanks{Corresponding author}\email{zhangxin@mail.neu.edu.cn}
\affiliation{Department of Physics, College of Sciences, Northeastern University, Shenyang 110819, China}
\affiliation{Key Laboratory of Data Analytics and Optimization for Smart Industry (Ministry of Education), Northeastern University, Shenyang 110819, China}
\affiliation{National Frontiers Science Center for Industrial Intelligence and Systems Optimization, Northeastern University, Shenyang 110819, China}

\begin{abstract}
Recent studies suggest that dark energy may be dynamical rather than being a mere cosmological constant $\Lambda$. In this work, we examine the viability of two physically well-motivated dynamical dark energy models---holographic dark energy (HDE) and Ricci dark energy (RDE)---by confronting them with the latest observational data, including ACT cosmic microwave background anisotropies, DESI baryon acoustic oscillations, and DESY5 supernovae. Our analysis reveals a fundamental tension between early- and late-universe constraints within both frameworks: ACT favors a quintom scenario where the dark energy equation of state evolves from $w>-1$ at early times to $w<-1$ at late times, while DESI+DESY5 exhibits a distinct preference for quintessence where $w>-1$ across cosmic evolution. Critically, the RDE model fails to provide a coherent description of cosmic evolution, as it manifests severe tensions (exceeding $10\sigma$ significance)  between early- and late-universe parameter reconstructions. In addition, Bayesian evidence disfavors both models relative to the $\Lambda$CDM model. Our findings statistically exclude the original HDE and RDE models and uncover a severe discrepancy between early- and late-universe observations described by them, leading to the conclusion that the HDE and RDE models can be ruled out by current observational data.
\end{abstract}

\maketitle
\section{Introduction}\label{sec1}
The observed accelerated expansion of the universe has led to the postulation of dark energy, a fundamental constituent accounting for $\sim70\%$ of the universe's energy density and characterized by its intrinsic negative pressure. Initially introduced by Einstein within the framework of General Relativity, the cosmological constant $\Lambda$ has re-emerged as the dominant theoretical paradigm for dark energy following the definitive confirmation of cosmic acceleration in 1998 \cite{SupernovaSearchTeam:1998fmf,SupernovaCosmologyProject:1998vns}. The $\Lambda$CDM model, which synthesizes $\Lambda$ (representing vacuum energy density), cold dark matter, and baryonic physics, stands as the prevailing cosmological standard model by virtue of its remarkable consistency with high-precision observational data, notably the cosmic microwave background (CMB) power spectrum measurements \cite{Aghanim:2018eyx}.

Notwithstanding its ubiquitous acceptance, the $\Lambda$CDM model is beset by persistent theoretical conundrums and observational discrepancies. Foremost among these is the cosmological constant problem: the striking disparity between quantum field theory (QFT) predictions for vacuum energy density and empirical estimates of $\Lambda$, with the former exceeding the latter by $\sim120$ orders of magnitude \cite{Weinberg:1988cp,Sahni:1999gb}. This ``\textit{worst prediction in physics}'' implies either a profound misunderstanding of vacuum energy in QFT or the existence of an unexplained fine-tuning mechanism cancelling 120 digits --- a scenario widely regarded as unnatural. Furthermore, the model fails to reconcile certain early- and late-universe observations, most notably in the $\sim5\sigma$ tension between the Planck CMB-based estimate of the Hubble constant $H_0=67.4\pm0.5\ {\rm km/s/Mpc}$ (assuming $\Lambda$CDM) and the  late-universe measurement of $H_0=73.04\pm1.04\ {\rm km/s/Mpc}$ from SH0ES \cite{Verde:2019ivm, Riess:2021jrx}.

The theoretical shortcomings and accumulating observational tensions confronting $\Lambda$CDM have spurred investigations into alternative cosmological frameworks, with particular emphasis on the dynamical nature of dark energy. For instance, spatially homogeneous scalar fields exhibiting slow-roll evolution can naturally generate the requisite negative pressure to drive the current phase of cosmic acceleration, thereby offering a compelling mechanism for dynamical dark energy \cite{Zlatev:1998tr,Steinhardt:1999nw,Peebles:1987ek,Ratra:1987rm}. Through precision constraints on the equation of state (EoS) parameter $w$ within such frameworks, one may discern potential deviations from $\Lambda$CDM's rigid $w=-1$ prescription. The phenomenological parameterizations, most notably the Chevallier-Polarski-Linder ansatz $w(a)=w_0+w_a(1-a)$ where $a$ is the scale factor \cite{Chevallier:2000qy,Linder:2002et}, are widely used for consistency tests of the $\Lambda$CDM model. For other parameterizations, see e.g. Refs.~\cite{Yang:2018qmz,Barboza:2008rh,Jassal:2004ej,Ma:2011nc}. It must be noted that such empirical constructions lack a theoretical foundation and thus provide little physical insight. Therefore, it becomes imperative to explore theoretically well-motivated dark energy scenarios.

The holographic dark energy (HDE) model is noted for its deep theoretical considerations \cite{Li:2004rb}. Rooted in the holographic principle, a cornerstone of modern theoretical physics first formalized by 't\,Hooft and Susskind, HDE posits that the dark energy density ($\rho_{\rm de}$) is intrinsically bounded by the entropy-area relation of a cosmic horizon. Specifically, the total energy in a region of scale $L$ should not exceed the mass of a black hole of the same size, $L^3\rho_{\rm de} \leq L M_{\rm p}^2$, where $M_{\rm p}$ is the reduced Planck mass \cite{Cohen:1998zx}. According to this bound, the density of HDE can be written as
\begin{equation}\label{HDE_density}
\rho_{\rm de} = 3c^2M_{\rm p}^2 / L^2,
\end{equation}
where $L$ is the infrared cutoff scale (typically chosen as the future event horizon), and $c$ is a dimensionless parameter which critically determines the properties of HDE. HDE differs from the cosmological constant paradigm by dynamically linking dark energy to spacetime's structure through ultraviolet/infrared duality. The significance of HDE lies in its capacity to resolve two long-standing paradoxes: (\romannumeral1) The cosmological constant problem is naturally mitigated through entropy-area scaling laws; (ii) The cosmic coincidence problem---why dark energy and dark matter densities are comparable today---finds a dynamical explanation via the evolving horizon scale \cite{Li:2004rb}.

A theoretical variant of HDE, the Ricci dark energy (RDE) model \cite{Gao:2007ep}, has also attracted significant interest. The distinction between HDE and RDE arises from their fundamental choice of infrared cutoff scale. HDE adopts the future event horizon radius, whereas RDE utilizes the characteristic length scale derived from the Ricci scalar curvature to define the dark energy density. Specifically,
\begin{align}\label{RDE_density}
\rho_{\rm de} = 3\gamma M_{\rm p}^2 (\dot{H} + 2H^2),
\end{align}
where $H$ is the Hubble parameter and the dot denotes the derivative with respect to cosmic time, $\gamma$ is a constant which plays an important role in determining the properties of RDE. Grounded in spacetime geometry itself, this model may offer insights into the interplay between quantum mechanics and General Relativity, which is pivotal for understanding the universe at its most fundamental level. Furthermore, the RDE model avoids the causality problem inherent in the HDE model through its foundation on the local spacetime geometry \cite{Gao:2007ep}. Therefore, RDE is also an attractive model for exploration.

The HDE and RDE models not only present a distinct perspective on the dynamical nature of dark energy but also demonstrate potential to resolve the puzzles faced by the $\Lambda$CDM model. Investigating them is essential for exploring quantum gravity-based explanations of dark energy. These two models have been constrained by astronomical probes, primarily the CMB anisotropies, baryon acoustic oscillations (BAO), and type Ia supernovae (SNe), see Refs.~\cite{Wang:2004nqa,Huang:2004wt,Nojiri:2005pu,Zhang:2005kj,Zhang:2005hs,Zhang:2006av,Zhang:2006qu,Setare:2006yj,Zhang:2007sh,Zhang:2007an,Zhang:2009un,Li:2009bn,Li:2010xjz,Zhang:2014ija,Landim:2015hqa,Feng:2016djj,Li:2012spm,Wang:2016och,Moradpour:2020dfm,Drepanou:2021jiv,Kibe:2021gtw,Zhang:2012uu,Zhang:2015rha,Colgain:2021beg} for earlier studies and Refs.~\cite{Wang:2023gov,Li:2024qus,Li:2024hrv,Tang:2024gtq,Trivedi:2024dju,Guin:2025xki} for recent progress. Despite extensive study, the models face persistent challenges, failing to match historical observational data \cite{Li:2010xjz,Xu:2016grp,Wen:2017aaa,Li:2024bwr}. They merit a renewed investigation with the latest observational data. Moreover, it must be emphasized that the majority of existing studies on these two models rely on combined early- and late-universe datasets, without exploring the potential observational discrepancies between different datasets. This constitutes a critical caveat. For instance, as noted in Ref.~\cite{Wu:2025wyk}, analyses utilizing late-universe observations appear to leave a room for the RDE model.

Recent observational data releases provide critical updates for cosmological constraints. For instance, the Atacama Cosmology Telescope (ACT) collaboration released the CMB measurements, based on their Data Release~6 with 5-year observations \cite{ACT:2025fju}. Around the same time, the Dark Energy Spectroscopic Instrument (DESI) collaboration reported the BAO measurements, based on their Data Release 2 \cite{DESI:2025zgx}. Earlier, the Dark Energy Survey (DES) program reported the high-quality SN samples discovered during its 5-year operation \cite{DES:2024tys}. These datasets hold great significance for measuring dark energy \cite{DESI:2025fii,DESI:2025gwf,Wu:2025wyk,Barua:2025ypw,Yang:2025oax,Scherer:2025esj,Silva:2025hxw,Li:2025dwz,Shah:2025ayl,Du:2025xes,Pang:2025lvh,Wu:2024faw,Li:2025owk,Paliathanasis:2025dcr,Hur:2025lqc,Lu:2025gki,Urena-Lopez:2025rad,Anchordoqui:2025fgz,Pan:2025psn,Pan:2025qwy,You:2025uon,Afroz:2025iwo,Dinda:2025iaq,Giare:2025pzu,Teixeira:2025czm,Ye:2025ulq,RoyChoudhury:2025dhe,Wolf:2025jed,Wolf:2024stt,Wang:2025zri,deSouza:2025rhv,Colgain:2025nzf,Gialamas:2024lyw,Cheng:2025lod,Luciano:2025elo,Cai:2025mas,Lee:2025hjw,Hussain:2025nqy,VanRaamsdonk:2025wvj,An:2025vfz,vanderWesthuizen:2025iam,Lin:2025gne,Tyagi:2025zov,Li:2024qso,Sabogal:2025jbo,Gialamas:2025pwv,Ozulker:2025ehg,Bayat:2025xfr,Hussain:2025vbo,Feleppa:2025clx,CosmoVerseNetwork:2025alb,Wu:2025vrl}. In this paper, we adopt these datasets to constrain the HDE and RDE models and conduct a cosmological analysis. This study rigorously assesses: (i) the evolutionary patterns of dark energy within these two frameworks; (ii) the consistency of parameters derived from early- and late-universe datasets;  (iii) whether these frameworks are supported by the current data compared to the standard $\Lambda$CDM model.

The paper is structured as follows: Section~\ref{sec2} presents the methodology, Section~\ref{sec3} details the observational data, Section~\ref{sec4} analyzes the constraints on HDE and RDE, and Section~\ref{sec5} concludes with key findings and their cosmological implications.

\section{methodology}\label{sec2}

We briefly review the HDE and RDE models. In HDE, the dark energy density is obtained by adopting the future event horizon as the infrared cutoff scale $L$ \cite{Li:2004rb}, defined by the integral:
\begin{align}
R_{\text{h}}(a) = a \int_{a}^{\infty} \frac{{\rm d}a'}{H(a') a'^2},
\end{align}
where $a=1/(1+z)$ is the scale factor. In a spatially flat FLRW universe, the Friedmann equation is given by
\begin{align}
3M_{\rm p}^2H^2= \rho_{\text{de}} + \rho_{\rm m} + \rho_{\rm r},
\end{align}
where $\rho_{\text{de}}$, $\rho_{\rm m}$ and $\rho_{\rm r}$ denotes the energy densities of dark energy, pressureless matter and radiation, respectively. The equation can be recast into the dimensionless Hubble parameter form:
\begin{align}
\label{Ez}
E(z)=H(z)/H_0=\sqrt{ \frac{ \Omega _{\rm r} (1+z)^4 + \Omega_{\rm m} (1+z)^3 }{ 1 - \Omega_{\rm de} (z) } },
\end{align}
where $\Omega_{\rm m}$ and $\Omega_{\rm r}$ are current fraction densities of matter and radiation, and $\Omega_{\text{de}}(z)$ is the fractional density of dark energy at redshift $z$ whose derivative is calculated by
\begin{align}
\label{derivative}
\frac{{\rm d}\Omega_{\text{de}}(z)}{{\rm d}z} &= -\frac{2\Omega_{\text{de}}(z)[1 - \Omega_{\text{de}}(z)]}{1 + z} \nonumber \\
& \times  \left( \frac{1}{2} + \frac{\sqrt{\Omega_{\text{de}}(z)}}{c}  + \frac{\sqrt{\Omega_{\text{r}}(z)}}{2[1 - \Omega_{\text{de}}(z)]}  \right).
\end{align}
By solving Eq.~(\ref{derivative}) and propagating its solution into Eq.~(\ref{Ez}), one can determine $E(z)$. The measurement of cosmological distances allows us to constrain the parameters in the $E(z)$ formalism. Given the energy conservation equation
\begin{align}
\label{conservation}
\dot{\rho}_{\rm de} + 3H\rho_{\rm de}(1 + w) = 0,
\end{align}
the dark energy EoS can be solved in the form
\begin{align}
\label{HDEEoS}
w(z)=-\frac{1}{3}- \frac{2\sqrt{\Omega_{\text{de}}(z)}}{3c}.
\end{align}
The parameter $c$ plays a crucial role in shaping the behavior of EoS. Dark energy models are typically classified into four main categories:

\begin{itemize}[noitemsep, topsep=0pt]
    \item \( w = -1 \): Represents the cosmological constant.
    \item \( w > -1 \): Corresponds to quintessence dark energy.
    \item \( w < -1 \): Refers to phantom dark energy.
    \item \( w \) crosses \( -1 \): Indicates quintom dark energy.
\end{itemize}
For the HDE model, when $c>1$, the EoS consistently stays above $-1$, displaying quintessence-like characteristics. However, when $c<1$, the EoS will cross the cosmological constant boundary at late times \cite{Zhang:2005hs}. The transition from $w>-1$ to $w<-1$ implies that the universe will be dominated by phantom-like dark energy \cite{Feng:2004ad}, which could lead to a catastrophic Big Rip, where the expansion of the universe becomes uncontrollably rapid, ultimately resulting in its demise.

The RDE model chooses the mean radius of the Ricci scalar curvature as the infrared cutoff to calculate the dark energy density. The corresponding Friedmann equation can be expressed in terms of the reduced Hubble parameter as
\begin{align}
E^{2} = \Omega_{\rm m} e^{-3x} + \gamma \left( \frac{1}{2} \frac{{\rm d}E^{2}}{{\rm d}x} + 2E^{2} \right),
\end{align}
where $x\equiv\ln a$. Solving this equation, we obtain
\begin{align}
\label{RDEEz}
E(z) =& \left[ \frac{2\Omega_{\rm m}}{2-\gamma}(1+z)^{3} +\Omega_{\rm r}(1+z)^{4}\right. \nonumber \\
 &\left.+ \left(1-\Omega_{\rm r}-\frac{2\Omega_{\rm m}}{2-\gamma}\right)(1+z)^{\left(4-\frac{2}{\gamma}\right)} \right]^{1/2}.
\end{align}
From Eq.~(\ref{RDEEz}), one can derive the fractional RDE density
\begin{align}
\label{RDEEoS}
\Omega_{\rm de}(z) =& \frac{\gamma}{2-\gamma}\Omega_{\rm m}(1+z)^3\nonumber \\
&+\left(1-\Omega_{\rm r}-\frac{2\Omega_{\rm m}}{2-\gamma}\right)(1+z)^{\left(4-\frac{2}{\gamma}\right)}.
\end{align}
The dark energy EoS satisfies
\begin{align}
\label{RDEEoS}
w(z) =-1 + \frac{1+z}{3}\frac{{\rm d}\ln{\Omega_{\rm de}}}{{\rm d}z}.
\end{align}
The parameter $\gamma$  is  pivotal in governing the dynamical behavior of RDE. Specific values of $\gamma$ lead to distinct cosmological implications: $\gamma>0.5$ corresponds to $w >-1$, while $\gamma<0.5$ leads to a phantom regime, i.e., RDE evolves from quintessence-like at early times to phantom-like at late times.

\section{Observational data}\label{sec3}

\begin{itemize}[leftmargin=1em, label=\textbullet]
\setlist[itemize,2]{label={\raisebox{0.5ex}{\tiny\rule{0.8em}{0.8pt}}}, leftmargin=1em}
    \item \textbf{Cosmic microwave background.}
    \begin{itemize}
    \item \textbf{ACT}: We employ the CMB data from ACT, which offers high-resolution observations of the CMB temperature, polarization, and cross spectra over $\sim40\%$ of the low-foreground sky \cite{ACT:2025fju}. Specifically, we adopt the ACT DR6 likelihood for temperature (TT), polarization (EE), and cross (TE), as well as the CMB lensing. This dataset is referred to as ACT.
    \item \textbf{Planck}: For a comparative analysis, the Planck data are utilized, which include the high-precision, full-sky measurements of the CMB TT, TE, and EE power spectra \cite{Aghanim:2018eyx,Efstathiou:2019mdh,Rosenberg:2022sdy}. This dataset is referred to as Planck.
    \end{itemize}

    \item \textbf{Baryon acoustic oscillations.} We adopt the BAO data from DESI DR2, based on the precise observations of bright galaxy sample (BGS), luminous red galaxies (LRG), emission line galaxies (ELGs), quasars (QSOs) and Lyman-$\alpha$ forests \cite{DESI:2025zgx}. Specially, we consider 13 BAO measurements including the BGS, LRG1, LRG2, LRG3+ELG1, ELG2, QSO and Lyman-$\alpha$ samples at the effective redshifts $z_{\rm eff} = 0.295$, $0.51$, $0.706$, $0.934$, $1.321$, $1.484$, and $2.33$, respectively. This dataset is referred to as DESI.

    \item \textbf{Type Ia supernovae.} We consider the SN data from DESY5, based on the five-year operation of DES program. The DESY5 sample consists 1829 distant SNe~Ia spanning $0.025<z<1.3$ \cite{DES:2024tys}. Compared to the PantheonPlus sample, the dataset quintuples the number of SNe beyond $z > 0.5$. This dataset is referred to as DESY5.
\end{itemize}

\section{Cosmological constraints}\label{sec4}

\begin{table*}[htb]
\renewcommand\arraystretch{1.5}
\centering
\normalsize
\caption{Fitting results ($1\sigma$ level) in the $\Lambda$CDM, HDE, and RDE models from the Planck, ACT, DESI+DESY5, and ACT+DESI+DESY5 data. Here, $H_{0}$ is in units of ${\rm km/s/Mpc}$. Note: Entries marked with ``--'' indicate parameters that are not effectively constrained by the data (i.e., posteriors are prior-dominated and lack physical meaning).}
\label{tab1}
\begin{tabular}{lccccc} 
\hline
Model&Parameter & Planck& ACT & DESI+DESY5 & ACT+DESI+DESY5 \\
\hline
$\Lambda$CDM	
& $H_{0}$                               & $67.21\pm 0.45$               & $66.89\pm 0.59$               & --                                  & $68.25\pm 0.29$ \\
& $\Omega_{\mathrm{m}}$     & $0.3160\pm 0.0060$              & $0.3239^{+0.0086}_{-0.0096}$  & $0.3104\pm 0.0078$                    & $0.3035\pm 0.0038$\\
& $\sigma_8$                        & $0.8114\pm 0.0046$            & $0.8177\pm 0.0049$            & --                 & $0.8119\pm 0.0044$\\

HDE
& $H_{0}$                   & $76.80^{+7.90}_{-4.40}$       & $67.00\pm 6.10$               & --                 & $67.32\pm 0.54$\\
& $\Omega_{\mathrm{m}}$     & $0.247^{+0.022}_{-0.052}$    & $0.336^{+0.054}_{-0.077}$    & $0.2723\pm 0.0087$                    & $0.3019\pm 0.0049$\\
& $\sigma_8$                & $0.890^{+0.062}_{-0.035}$     & $0.821^{+0.057}_{-0.066}$     & --                 & $0.7701\pm 0.0086$\\
& $c$                       & $0.458^{+0.033}_{-0.096}$     & $0.649^{+0.084}_{-0.240}$     & $1.086^{+0.073}_{-0.093}$             & $ 0.725^{+0.022}_{-0.025}$\\

RDE
& $H_{0}$                   & $82.51^{+0.49}_{-0.27}$       & $ 84.60^{+1.70}_{-0.57} $      & --                & $ 74.55\pm 0.54$ \\
& $\Omega_{\mathrm{m}}$     & $0.2528^{+0.0023}_{-0.0036}$  & $0.2520^{+0.0036}_{-0.0110}$  & $0.2162^{+0.0082}_{-0.0072}$          & $0.2289\pm 0.0035$ \\
& $\sigma_8$                & $0.9224\pm 0.0083$            & $0.962^{+0.017}_{-0.013}$     & --              & $ 0.409\pm 0.011$\\
& $\gamma$                  & $0.1440\pm 0.0031$            & $0.1324\pm 0.0039$            & $0.552^{+0.015}_{-0.018}$          & $0.3649\pm 0.0062$  \\

\hline
\end{tabular}
\end{table*}

\begin{figure*}
\includegraphics[scale=0.57]{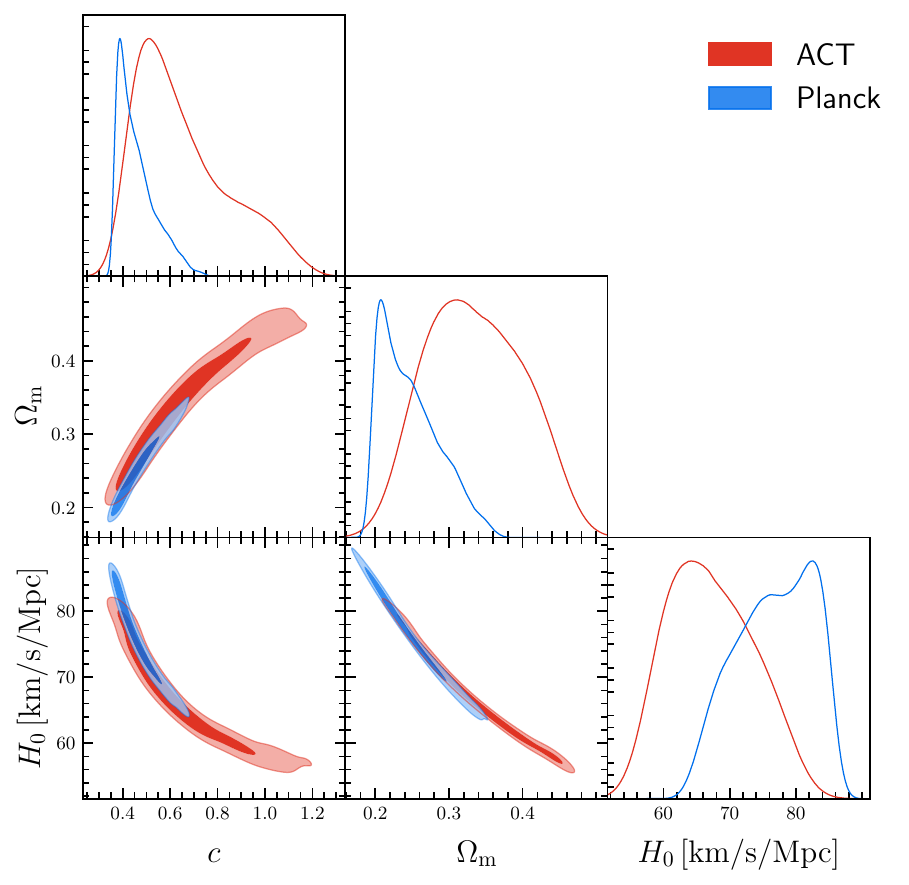}
\includegraphics[scale=0.57]{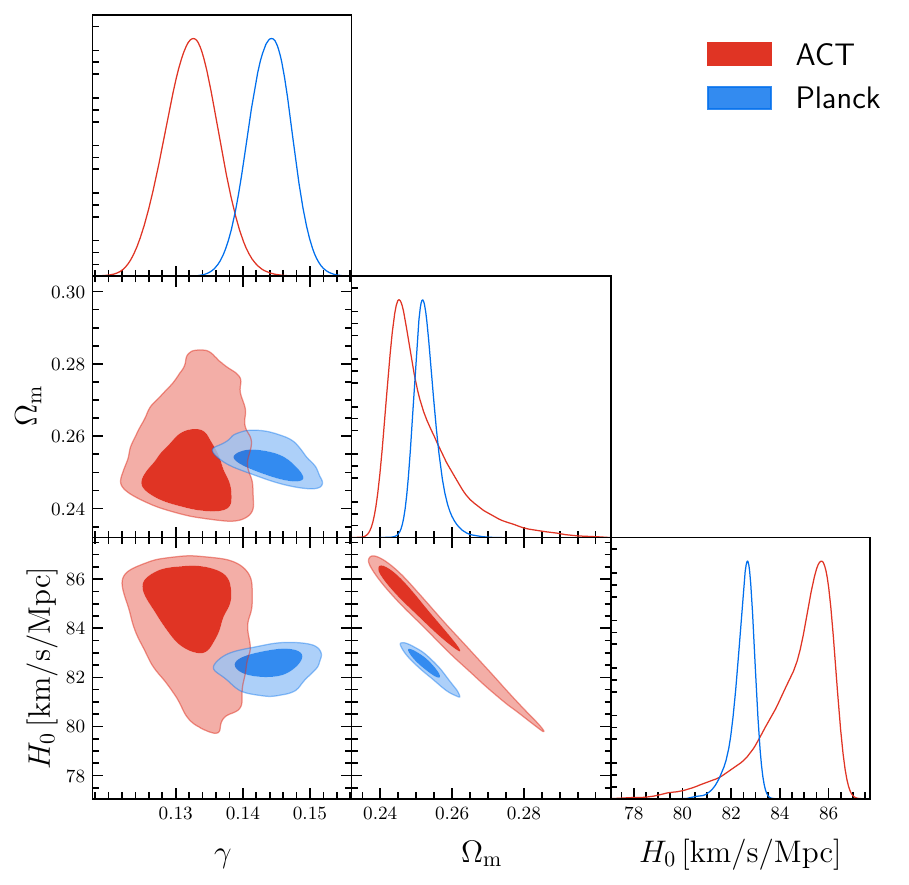}
\centering
\caption{Constraints on cosmological parameters using the Planck and ACT CMB data. Left panel: Constraints on the HDE model. Right panel: Constraints on the RDE model.}
\label{PlanckACT}
\end{figure*}

\begin{figure*}
\includegraphics[scale=0.57]{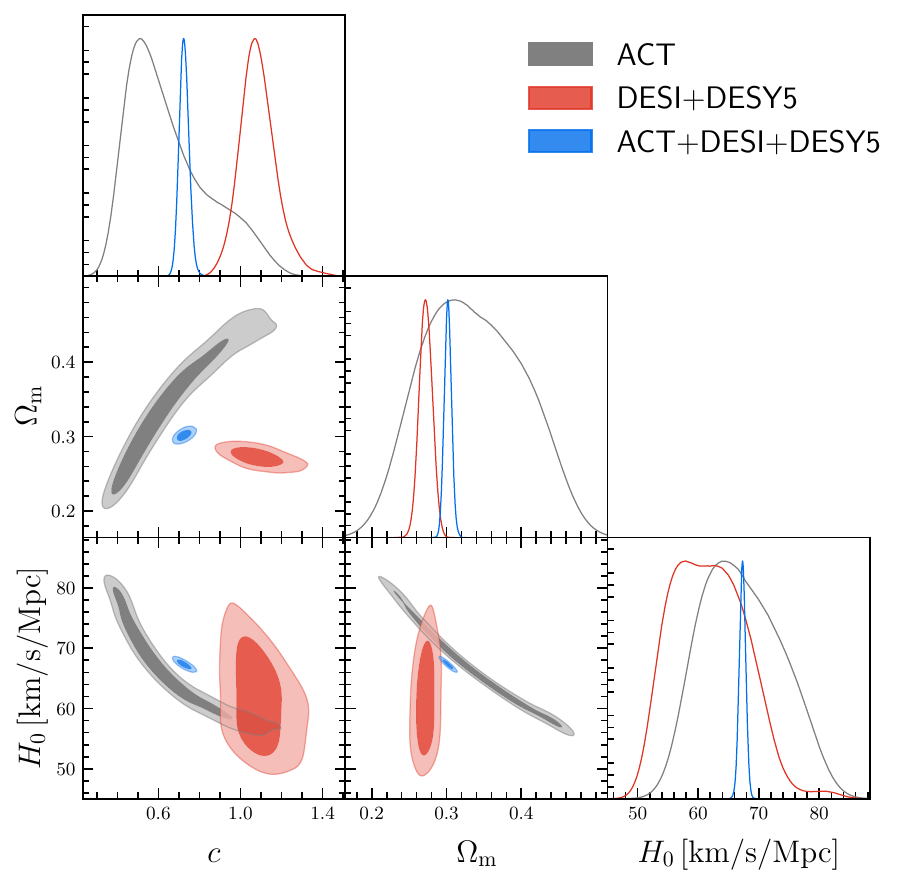}
\includegraphics[scale=0.57]{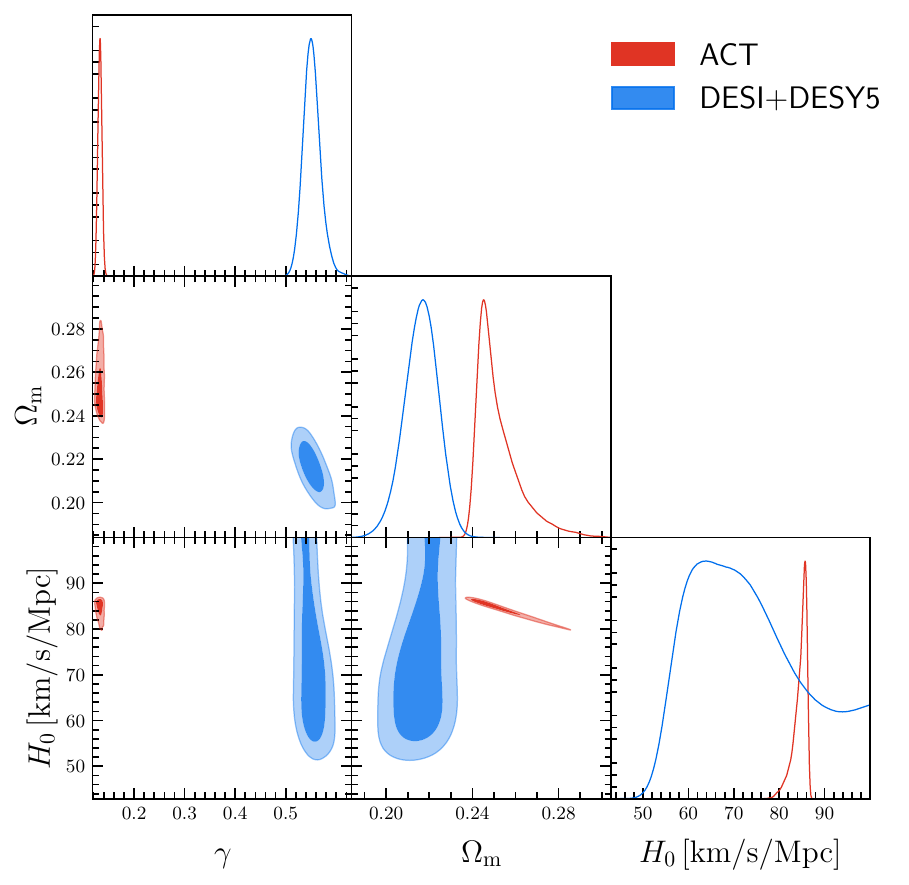}
\centering
\caption{Constraints on cosmological parameters using the ACT, DESI+DESY5, and ACT+DESI+DESY5 data. Left panel: Constraints on the HDE model. Right panel: Constraints on the RDE model.}
\label{CBS}
\end{figure*}

In this section, we adopt the Markov Chain Monte Carlo method to infer the probability distributions of cosmological parameters using the observational data, by maximizing the likelihood $L\propto\rm{exp}(-\chi^2/2)$. The $\chi^2$ function for BAO and SN data can be written as $\Delta\boldsymbol{D}^T\boldsymbol{C}^{-1}\Delta\boldsymbol{D}$, where $\Delta\boldsymbol{D}$ is the vector of observable residuals representing the difference between observation and theory, and $\boldsymbol{C}$ is the covariance matrix. We compute theoretical models with the {\tt CAMB} code \cite{Gelman:1992zz} and conduct parameter sampling with the publicly available {\tt Cobaya} \cite{Torrado:2020dgo}. The chain convergence is assessed via the Gelman-Rubin potential scale reduction factor ($R-1<0.02$ for all parameters). We fit the standard $\Lambda$CDM (comparison baseline), HDE, and RDE models to multiple data combinations. The $1\sigma$ and $2\sigma$ parameter distribution contours are shown in Figs.~\ref{PlanckACT} and \ref{CBS}, and the $1\sigma$ errors for the marginalized parameter constraints are summarized in  Table~\ref{tab1}. It is essential to emphasize that the constraints on the Hubble constant $H_0$ and the matter fluctuation amplitude $\sigma_8$ derived from late-time geometric probes are fundamentally limited. For $H_0$, this limitation arises from the strong degeneracy with the sound horizon at the drag epoch $r_{\rm d}$; for $\sigma_8$, it stems from the lack of cosmic structure growth data. The physical scale of $r_{\rm d}$ is determined by early-universe physics, and without an external anchor---typically provided by Big Bang Nucleosynthesis (BBN) or CMB measurements---the combination of BAO and SNe cannot break the  $H_0-r_{\rm d}$ degeneracy. Likewise, $\sigma_8$ cannot be constrained without growth-sensitive probes such as the redshift-space distortions. We mark the corresponding entries in tables with ``--'' to indicate that they are not effectively constrained.

We begin by examining constraints derived from two CMB datasets. Among these, Planck's full-sky coverage delivers unparalleled precision in measuring large-scale anisotropies, thereby minimizing cosmic variance. Additionally, its broader spectral coverage enables superior foreground removal. In contrast, ACT conducts deep observations over smaller patches of low-foreground sky, achieving higher angular resolution at small scales. However, its limited sky coverage inherently restricts the statistical power for constraining cosmological parameters that sensitive to large angular scales. Consequently, the Planck data yield stronger constraining power than that from ACT (as shown in Fig.~\ref{PlanckACT}), aligning with prior studies \cite{ACT:2025fju,Ferreira:2025lrd,DESI:2025gwf,ACT:2025tim}. Notably, both Planck and ACT impose tighter constraints on the RDE model than on the HDE model, as demonstrated in Table~\ref{tab1} and Fig.~\ref{PlanckACT}. This difference arises because RDE's energy density depends directly on the spacetime curvature, to which CMB anisotropies are highly sensitive, whereas HDE's horizon-based scale involves integrated dynamical effects, making its imprints more degenerate with other cosmological parameters. In the subsequent analysis, we discuss the constraints on HDE and RDE from both Planck and ACT.

In the HDE model, Planck and ACT yield the Hubble constant estimates of $H_0=76.80^{+7.90}_{-4.40}\ {\rm km/s/Mpc}$ and $H_0=67.00\pm 6.10\ {\rm km/s/Mpc}$ respectively. The large uncertainties ($\sim10\%$) in these constraints make them statistically consistent with the local distance-ladder measurement by the SH0ES collaboration ($H_0=73.04\pm1.04\ {\rm km/s/Mpc}$). In contrast, CMB observations place tight constraints on $H_0$ in the RDE model. Specifically, Planck yields $H_0=82.51^{+0.49}_{-0.27}\ {\rm km/s/Mpc}$, while ACT gives $H_0=84.60^{+1.70}_{-0.57}\ {\rm km/s/Mpc}$. In stark contrast to the SH0ES result, both values are significantly elevated, resulting in a statistical tension of $8.2\sigma$ for Planck and $5.8\sigma$ for ACT. These results indicate that RDE does not alleviate the Hubble tension;  on the contrary, its dynamical properties further exacerbate the discrepancy.

An anti-correlation between the matter density parameter $\Omega_{\rm m}$ and the matter fluctuation amplitude $\sigma_8$ is observed, reflecting the inherent geometric degeneracies in CMB power spectrum analysis. The Planck data indicate a preference for a lower matter density parameter $\Omega_{\rm m}$ and a higher amplitude of matter fluctuations $\sigma_8$ in both HDE and RDE models, relative to the values inferred within the $\Lambda$CDM framework. In particular, HDE yields $\Omega_{\rm m}=0.247^{+0.022}_{-0.052}$ and $\sigma_8=0.890^{+0.062}_{-0.035}$, while RDE gives $\Omega_{\rm m}=0.2528^{+0.0023}_{-0.0036}$ and $\sigma_8=0.9224\pm 0.0083$. In addition, ACT observations favor a similar trend for the RDE model ($\Omega_{\rm m}=0.2520^{+0.0036}_{-0.0110}$ and $\sigma_8=0.962^{+0.017}_{-0.013}$), but reveal a slight departure in the HDE model: $\Omega_{\rm m}=0.336^{+0.054}_{-0.077}$ and $\sigma_8=0.821^{+0.057}_{-0.066}$. In the HDE model, both Planck and ACT results remain consistent with late-universe DESI+DESY5 constraints on $\Omega_{\rm m}$, showing only marginal deviations ($<1.2\sigma$). In contrast, CMB results exhibit severe tensions with late-universe constraints in the RDE model. Specifically, the DESI+DESY5 result ($\Omega_{\rm m}=0.2162^{+0.0082}_{-0.0072}$) exhibits a $4.8\sigma$ tension with the Planck result and a $4.4\sigma$ tension with the ACT result.

The dark energy parameters ($c$ for HDE and $\gamma$ for RDE) exhibit notable discrepancies across CMB datasets. In particular, Planck yields a lower value of $c=0.458^{+0.033}_{-0.096}$ compared to the ACT value, $c=0.649^{+0.084}_{-0.240}$, corresponding to a tension at the $1.5\sigma$ level. In the RDE model, Planck favors a higher $\gamma$ value ($\gamma=0.1440\pm 0.0031$) than ACT ($\gamma=0.1324\pm 0.0039$), leading to a $2.3\sigma$ tension. Despite these inconsistencies, both CMB datasets consistently prefer $c<1$ for the HDE model and $\gamma<0.5$ for the RDE model, thus supporting a quintom scenario, i.e., dark energy evolves from a quintessence-like state in early times to a phantom-like regime at late times. In the following, we assess the consistency between dark energy parameters derived separately from early-universe (ACT) and late-universe (DESI+DESY5) datasets.

\begin{figure*}
\includegraphics[scale=0.35]{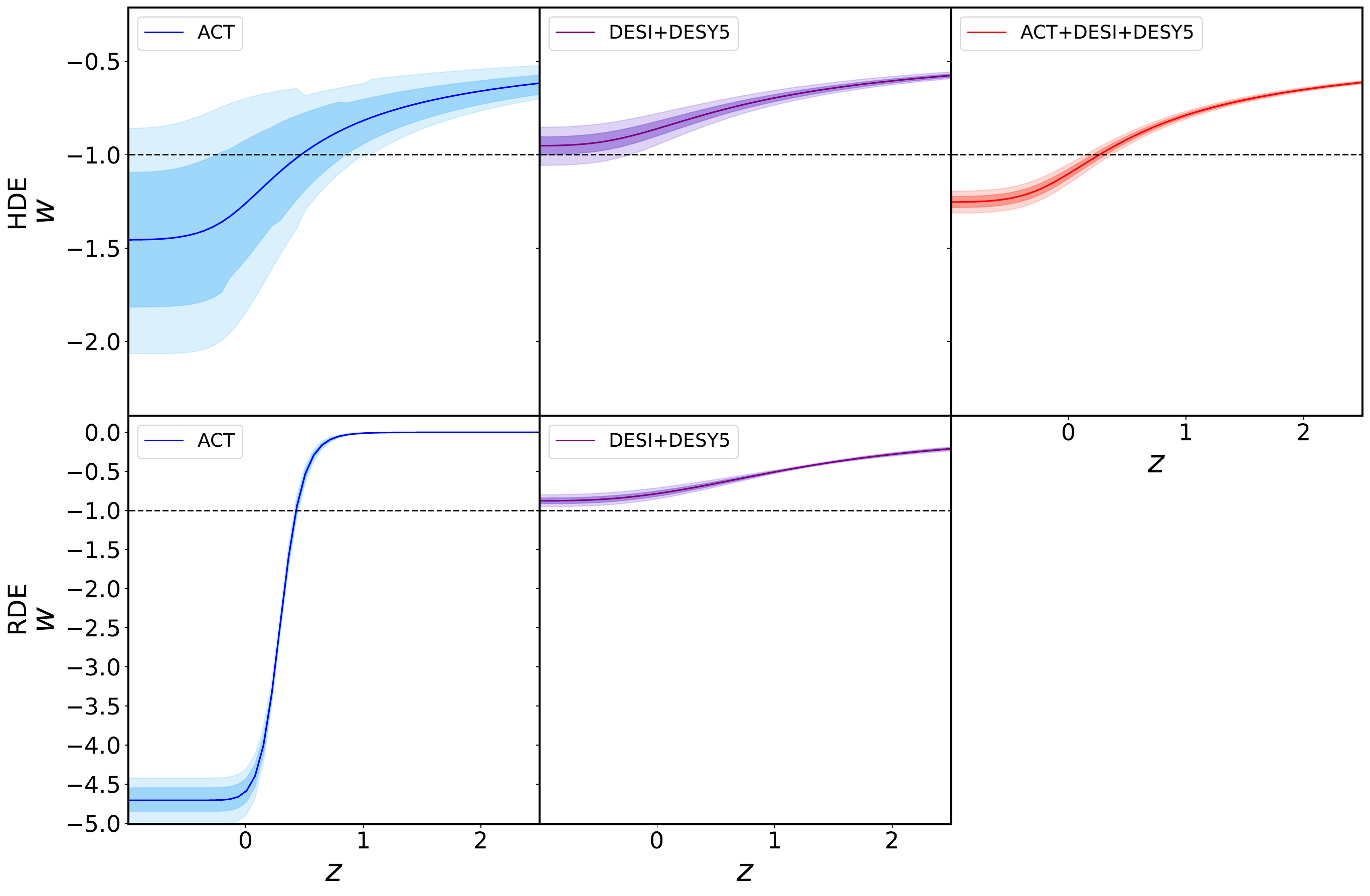}
\centering
\caption{The reconstructed evolution history of dark energy EoS $w$ at the $1\sigma$ and $2\sigma$ confidence levels for the HDE and RDE models, constrained by current observational data. The black dashed line denotes the cosmological constant  ($w=-1$).}
\label{HDE-RDE-EoS}
\end{figure*}

\begin{figure*}
\includegraphics[scale=0.4]{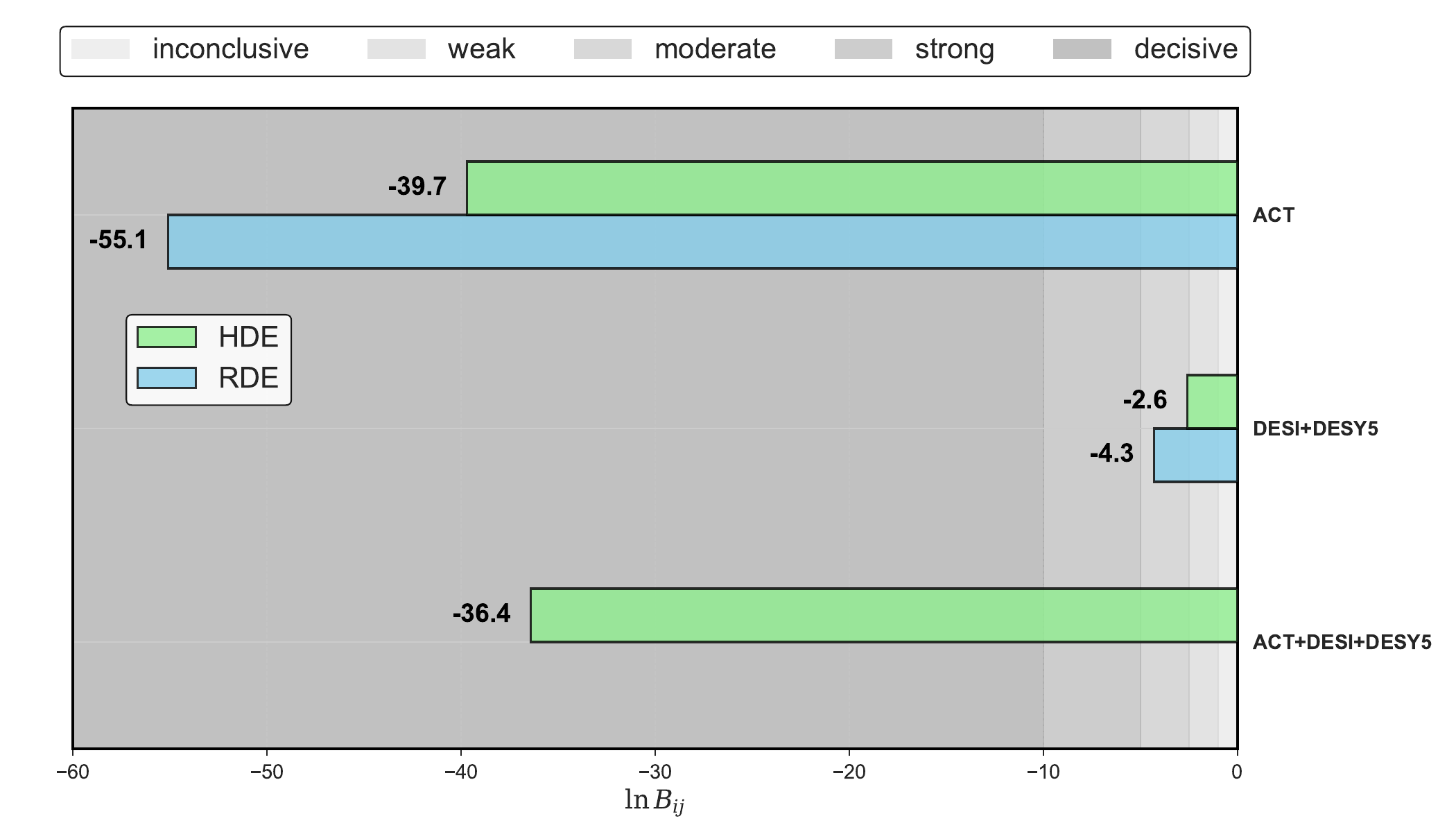}
\caption{Graphical representation of dark energy model comparison results. The Bayes factor $\ln \mathcal{B}_{ij}$ (where $i$ denotes alternative dark energy models and $j$ the $\Lambda$CDM model) quantifies relative model evidence according to Jeffreys scale, with negative values signifying statistical preference for $\Lambda$CDM.  The shaded regions from left to right represent inconclusive, weak, moderate, strong, and decisive evidence, respectively.}
\label{Bayes-factor}
\end{figure*}

As expressed by Eq.~(\ref{HDEEoS}), the parameter $c$ critically regulates the HDE EoS. For cases with $c>1$, $w>-1$ holds universally, indicating quintessence-like behavior and the universe experiences a gradually weakening acceleration; when $c=1$, $w$ evolves from values greater than $-1$ at early times to asymptotically approach $-1$ in late epochs, mimicking a cosmological constant and resulting in a stable de Sitter future; if $c<1$, $w$ will drop below $-1$ in late epochs, signifying phantom-like dynamics that may lead to a Big Rip singularity. The joint data yield $c=0.725^{+0.022}_{-0.025}$ for the HDE model. Notably, we observe a $1.7\sigma$ tension between the ACT and DESI+DESY5 datasets: the former prefers $c=0.649^{+0.084}_{-0.240}$, whereas the latter gives $c=1.086^{+0.073}_{-0.093}$. Although this discrepancy remains below the $3\sigma$ threshold, a more significant tension emerges when comparing the DESI+DESY5 value with the Planck result, reaching the $5.2\sigma$ level. These results indicate that the HDE model has difficulty in simultaneously accommodating observational data from both ends of cosmic history.

The parameter $\gamma$ also critically regulates the EoS of RDE as expressed by Eq.~(\ref{RDEEoS}).  For instance, $\gamma=0.5$ recovers the cosmological constant, $\gamma >0.5$ yields quintessence-like behavior, and $\gamma<0.5$ leads to a quintom regime.
We find a marked discrepancy emerges between early- and late-universe probes: the ACT data alone constrain $\gamma$ to $\gamma=0.1324\pm0.0039$, which stands in stark tension (exceeding $20\sigma$) with the value derived from DESI+DESY5, $\gamma=0.552^{+0.015}_{-0.018}$. The tension also exceeds $20\sigma$ when comparing Planck value ($\gamma=0.1440\pm0.0031$) with the DESI+DESY5 result. Note that there are better methods to quantify measurement inconsistencies, such as parameter-shift, Suspiciousness, and posterior predictive tests. However, to maintain consistency with the approach adopted throughout this work, we have continued to use the simpler tension estimate. Given the irreconcilable discrepancies in the estimation of key parameters, we conclude that the canonical RDE model fails to provide a unified framework for cosmic evolution. This result is consistent with previous studies \cite{Xu:2016grp,Wen:2017aaa,Li:2024bwr}, which have largely relied on joint analyses of early- and late-universe observations. This work provides clear evidence that the RDE model exhibits severe tensions between early- and late-universe observations. Given these inconsistencies, a joint analysis is uninformative and therefore is not presented here.

We present the redshift evolution of dark energy EoS constrained under HDE and RDE frameworks in Fig.~\ref{HDE-RDE-EoS}. It is evident that, within both frameworks, ACT favors an EoS that evolves from quintessence-like at early times to phantom-like at late times, whereas DESI+DESY5 exhibits a distinct preference for quintessence throughout cosmic evolution, although in HDE, this conclusion is only supported at approximately $1\sigma$ confidence. The joint analysis for HDE align more closely with the ACT preference, favoring an EoS crossing the phantom divide. When using the joint data, the transition from quintessence to phantom regime occurs within the redshift range $0<z<1$,  implying that the current dark energy EoS is less than $-1$, thereby producing a stronger negative pressure than that exerted by the cosmological constant $\Lambda$. In summary, the reconstruction of $w(z)$ from early- and late-universe observations supports divergent evolutionary pathways, and hence distinct cosmic destinies. Since there can be only one physical universe, this persistent tension severely undermines the credibility of both HDE and RDE models.

To assess the two holographic-inspired dynamical dark energy scenarios, we compare them with the $\Lambda$CDM model in terms of their ability to fit the current observational data. As models become more complex (i.e., have more free parameters), they tend to fit the observational data better, which can lead to a lower $\chi^2$ value. Therefore, the $\chi^2$ comparison is unfair for comparing different models. In this work,  we employ Bayesian evidence as a quantitative measure of model performance. The evidence for a model $M$ is defined as
the marginal likelihood of the observational data $D$:
\begin{equation}
Z = \int_{ \Omega} P(D|{\bm \theta}, M) P({\bm \theta}|M) P(M){\rm d}{\bm \theta},
\end{equation}
where \( P(D|\boldsymbol{\theta}, M) \) is the likelihood of data given parameters $\bm \theta$ and model $M$, \( P(\boldsymbol{\theta}|M) \) is the probability of $\bm \theta$ given $M$, and $ P(M)$ is the prior of $M$ itself. The logarithmic Bayes factor comparing models \( i \) and \( j \) is
\begin{equation}
\ln {\mathcal{B}}_{ij} = \ln Z_i - \ln Z_j,
\end{equation}
with \( Z_i \) and \( Z_j \) being the evidence values for models \( i \) and \( j \) respectively. Compared to model $j$, model $i$ is regarded as inconclusive supported when $ 0<\ln \mathcal{B}_{ij} <1$, weakly supported when $1\leq \ln \mathcal{B}_{ij} <2.5$, moderately supported when $2.5\leq \ln \mathcal{B}_{ij} <5$ , strongly supported when $5\leq \ln \mathcal{B}_{ij} <10$, and decisively supported when $\ln \mathcal{B}_{ij} \geq10$. Conversely, a negative $ \ln \mathcal{B}_{ij}$ value is interpreted as evidence against the model $i$.

Fig.~\ref{Bayes-factor} visually presents the numerical result, where $i$ denotes the HDE/RDE model and $j$ refers to the $\Lambda$CDM model. As can be seen, ACT decisively favors $\Lambda$CDM over both HDE and RDE scenarios, while DESI+DESY5 demonstrate only moderate statistical preference for the $\Lambda $CDM model. Specifically, ACT yields $\ln \mathcal{B}_{ij}=-39.7$ for HDE and $-55.1$ for RDE, and DESI+DESY5 provides $ \ln \mathcal{B}_{ij} =-2.6$ for HDE and $-4.3$ for RDE. The decisive disfavor of HDE and RDE by the CMB data appears linked to their altered expansion histories relative to $\Lambda$CDM. The locations and amplitudes of the CMB acoustic peaks are sensitive to the sound horizon at recombination and the subsequent expansion rate. Both HDE and RDE introduce dark energy dynamics that modify $H(z)$ at high redshifts, leading to a poor fit to the measured angular scale of sound horizon. In contrast, late-time measurements mainly capture the recent period of cosmic evolution, during which $H(z)$ of these models and that of $\Lambda$CDM differ less clearly. Additionally, adjustments to other parameters could improve the agreement with late-time geometry. The DESI+DESY5 results are consistent with the analysis reported in Ref.~\cite{Wu:2025wyk}. Any observationally viable model must be consistent with both early- and late-universe observations. Unfortunately, neither HDE nor RDE is able to satisfy this requirement. The joint analysis conclusively validates the $\Lambda$CDM framework while achieving high-significance exclusion of HDE alternatives.

\section{Conclusion}\label{sec5}
Previous studies have indicated that dark energy may exhibit dynamical evolution rather than being a mere cosmological constant. In this paper, we utilize the latest observational data, including CMB, BAO, and SN data, to examine the viability two well-motivated dynamical dark energy models, namely holographic dark energy and Ricci dark energy models. We constrain these two models separately using early-universe data, late-universe observations, and the data combination. Our primary aim is to explore the evolutionary patterns of dark energy within these cosmological frameworks and assess whether these frameworks are supported by the current observational data.

We constrain the two holographic-inspired models with the CMB data from the space-based Planck and ground-based ACT respectively, demonstrating that Planck imposes tighter constraints than ACT. Moreover, the two CMB experiments yield mutually consistent constraints on each model, with no statistically significant discrepancies observed. Due to their intrinsic properties, the CMB data provide high-precision constraints on the RDE model (achieving $\sim1\%$ parameter precision) but fail to impose tight constraints on the HDE model. The CMB data prefer $c<1$ for the HDE model and $\gamma<0.5$ for the RDE model, indicating a quintom dark energy where the equation of state parameter $w$ crosses the phantom divide at $w=-1$ in late epochs, evolving from a quintessence-like ($w>-1$) state to a phantom-like ($w<-1$) regime. The transition from $w>-1$ to $w<-1$ occurs between $0<z<1$, implying that the universe is currently dominated by the phantom-like dark energy in the HDE/RDE framework, which could lead to a Big Rip singularity in the far future.

Notably, a significant tension emerges between early-universe and late-universe constraints. The Planck and ACT data prefer values of $c<1$ and $\gamma<0.5$. In contrast, the DESI+DESY5 data favor $c>1$ for the HDE model and $\gamma>0.5$ for the RDE model, suggesting a quintessence-like evolutionary behavior wherein the universe maintains a gradually decelerating acceleration. For the HDE model, the tension in the parameter $c$ reaches $1.7\sigma$ between ACT and DESI+DESY5, and $5.2\sigma$ between Planck and DESI+DESY5. For the RDE model, the tension in the parameter $\gamma$ exceeds $20\sigma$ between ACT and DESI+DESY5, as well as between Planck and DESI+DESY5. Combined constraints yield parameter values consistent with $c<1$ and $\gamma<0.5$, specifically $c=0.725^{+0.022}_{-0.025}$ at the $1\sigma$ confidence level. It should be stressed that when discrepancies between different datasets exceed the $3\sigma$ threshold, a combined analysis is neither feasible nor physically justified. Therefore, the joint constraints should be interpreted with caution and regarded as nominal estimates rather than robust conclusions.

Within the RDE framework, profound tensions are observed not only in the parameter $\gamma$, but across the parameter space. Notably, the Hubble constant $H_0$ derived from the CMB data is in severe disagreement with the local distance-ladder measurement by the SH0ES team. This discrepancy reaches a statistical significance of $8.2\sigma$ between Planck and SH0ES, and $5.8\sigma$ between ACT and SH0ES. These results indicate that the RDE model provides no solution to the Hubble tension. In principle, a viable cosmological framework must demonstrate fundamental concordance between empirical constraints from the early- and late-universe. Such pronounced inconsistencies collectively cast substantial doubt on the robustness of the RDE model in describing cosmic evolution. In contrast, the HDE model exhibits less pronounced tensions in parameter constraints compared to the RDE model, though the inconsistencies remain non-negligible.

Bayesian evidence quantification disfavors both HDE and RDE models relative to the $\Lambda$CDM model. The conclusion is driven by decisive evidence from early-universe observations, supplemented by moderate disfavor from the late-universe data. These findings critically undermine the empirical viability of both RDE and HDE as physical descriptions of dark energy. Consequently, these models can be ruled out by the current data, and future investigations should accordingly prioritize alternative dark-energy scenarios. Of course, this criticism only applies to the specific versions of HDE and RDE tested in this paper, their extended theoretical frameworks could still be valid candidates.

\begin{acknowledgments}
This work was supported by the National SKA Program of China (Grants Nos. 2022SKA0110200 and 2022SKA0110203), the National Natural Science Foundation of China (Grants Nos. 12533001, 12575049, and 12473001), and the National 111 Project (Grant No. B16009).

\end{acknowledgments}

\bibliography{HDE+RDE}

\end{document}